\documentclass[prb,twocolumn]{revtex4}
\usepackage{amssymb}
\usepackage{amsmath}
\usepackage{epsfig}
\usepackage{bbm}
\usepackage{graphicx}

\newcommand{\mtin}[1]{\mbox{\tiny {#1}}}
\newcommand{\ca}[1]{{\cal #1}}
\newcommand{\ovl}[1]{\overline{#1}}
\newcommand{\sfrac}[2]{{\textstyle\frac{#1}{#2}}}

\def\bk{{\bf k}}

\def\bp{{\bf p}}
\def\bq{{\bf q}}

\def\bQ{{\bf Q}}

\def\b0{{\bf 0}}

\def\cS{{\cal S}}

\def\bra{\langle}
\def\ket{\rangle}
\def\up{\uparrow}
\def\down{\downarrow}

\def\eps{\epsilon}
\def\gam{\gamma}
\def\Gam{\Gamma}

\def\Lam{\Lambda}

\def\sg{\sigma}

\begin{document}

\bibliographystyle{apsrev}

\title{Incommensurate nematic fluctuations in the two-dimensional 
 Hubbard model}
\date{\today}
\author{Christoph Husemann}
\author{Walter Metzner}
\affiliation{Max Planck Institute for Solid State Research, 
 D-70569 Stuttgart, Germany}

\begin{abstract}
We analyze effective $d$-wave interactions in the two-dimensional 
extended Hubbard model at weak coupling and small to moderate doping.
The interactions are computed from a renormalization group flow.
Attractive $d$-wave interactions are generated via antiferromagnetic 
spin fluctuations in the pairing and charge channels.
Above Van Hove filling, the $d$-wave charge interaction is maximal
at incommensurate diagonal wave vectors, corresponding to nematic
fluctuations with a diagonal modulation.
Below Van Hove filling a modulation along the crystal axes can be 
favored. 
The nematic fluctuations are enhanced by the nearest-neighbor 
interaction in the extended Hubbard model, but they always remain
smaller than the dominant antiferromagnetic, pairing, or charge 
density wave fluctuations.
\end{abstract}

\maketitle

\section{Introduction}

Numerous layered correlated electron compounds exhibit competing 
instabilities, which are driven by a variety of dynamically 
generated effective interactions.
This feature is already borne out by the simplest model of a 
single layer in such systems, the two-dimensional Hubbard model.
Near half-filling, a whole zoo of enhanced fluctuations
is generated in that model, not only in the doped Mott insulator 
regime at strong coupling,\cite{lee06} but also for a weak bare 
interaction.\cite{metzner12,zanchi00,halboth00,honerkamp01,kampf03}
In addition to the pronounced antiferromagnetic and $d$-wave
superconducting fluctuations, an attractive charge forward 
scattering interaction with a $d_{x^2-y^2}$-wave symmetry was 
found.\cite{halboth00a,yamase00}
If strong enough, such an interaction can trigger a $d$-wave
Pomeranchuk instability leading to a nematic state with broken
orientation symmetry.\cite{fradkin10}

Nematic order with a $d_{x^2-y^2}$ symmetry has been observed in
several layered correlated electron compounds.
A nematic phase with a sharp phase boundary has been established
in a series of experiments on ultrapure $\rm Sr_3 Ru_2 O_7$
crystals in a strong magnetic field.\cite{ruthenate}
Electronic nematicity has also been observed in the high 
temperature superconductor $\rm Y Ba_2 Cu_3 O_y$ in transport 
experiments \cite{ando02,daou10} and neutron scattering.
\cite{hinkov}

Nematic tendencies compete with other instabilities.
In a functional renormalization group (fRG) analysis, with
forward scattering and other interaction channels treated on 
equal footing, nematic fluctuations in the two-dimensional 
Hubbard model were found to be weaker than antiferromagnetism 
and $d$-wave superconductivity.\cite{honerkamp02}
However, this does not necessarily prevent a nematic state,
since nematic order may coexist with antiferromagnetism 
\cite{kivelson98} and superconductivity.\cite{neumayr03,yamase07}
Moreover, the fRG calculations are reliable only in the weak 
coupling regime. A pronounced nematicity has been obtained by
dynamical cluster mean-field calculations for the
two-dimensional Hubbard model at strong coupling.
\cite{okamoto10} 

In a recent analysis of secondary instabilities generated by
antiferromagnetic fluctuations in a two-dimensional electron 
system with a nearly half-filled band, 
Metlitski and Sachdev \cite{metlitski10a,metlitski10b}
found a tendency toward formation of a {\em modulated} nematic
state. In that state the nematic order oscillates across the
crystal, with generally incommensurate wave vectors that point
along the Brillouin zone diagonal and connect antiferromagnetic 
hot spots with collinear Fermi velocities. 
At the same wave vectors the bare $d$-wave polarization function 
of tight-binding electrons on a square lattice has pronounced
peaks, which also indicates that a modulated nematic state is
favored over a homogeneous one.\cite{holder12}
Upon reducing the electron concentration the hot spots move 
toward the saddle points of the dispersion and the modulation 
vector shrinks, until it vanishes at Van Hove filling. 
Below Van Hove filling a modulation with a wave vector along 
the crystal axes can be favored.\cite{holder12}

In this article we analyze the tendency toward formation of a
modulated nematic state in the two-dimensional Hubbard model
and its extension with a nearest-neighbor interaction.
To this end we compute the effective two-particle interaction
from a fRG flow equation and compare the nematic channel to
antiferromagnetic, charge density, and pairing interactions.
All channels are treated on equal footing and the approximations
are controlled at weak coupling. 
We find that away from Van Hove filling a modulated nematic 
instability can indeed be favored over a homogeneous one, but
remains in any case subleading compared to antiferromagnetism
and $d$-wave superconductivity.
The nearest-neighbor interaction present in the extended
Hubbard model enhances the nematic fluctuations.

The paper is structured as follows.
In Sec.~II we introduce the extended Hubbard model and some
notation. 
In Sec.~III we describe the fRG flow equation and specify the 
parametrization of the effective two-particle interaction.
Results for the effective interaction are presented and
discussed in Sec.~IV.
A summary with the main conclusions follows in Sec.~V.



\section{Extended Hubbard model}

We will analyze the extended Hubbard model on a square lattice with 
a nearest neighbor hopping amplitude $t>0$ and a next-to-nearest 
neighbor hopping $t' < 0$.
In standard second quantization notation, the corresponding
Hamiltonian reads
\begin{equation} \label{hubbard}
 H = \sum_{\mathbf{k},\sigma} \epsilon_{\mathbf{k}} 
 a^{\dagger}_{\mathbf{k}\sigma} a_{\mathbf{k}\sigma} +  
 U \sum_j n_{j\up} n_{j\down} + 
 V \sum_{\bra j,j' \ket} n_j n_{j'} \, ,
\end{equation}
with a tight-binding dispersion of the form
$\eps_{\mathbf{k}} = 
 -2t (\cos k_x + \cos k_y) - 4t' \cos k_x \cos k_y$.
The Hubbard interaction $U>0$ is an on-site repulsion of electrons
with opposite spin.
The operator $n_{j\sg}$ counts electrons with spin orientation 
$\sg$ on the lattice site $j$, and $n_j = n_{j\up} + n_{j\down}$. 
The second interaction term with $V>0$ is a density-density repulsion 
between nearest neighbor lattice sites.
The lattice sum over nearest neighbors $\bra j,j' \ket$ is defined 
such that each bond is taken into account only once.
The nearest neighbor interaction favors charge order.
In mean field theory there is a first order transition between 
spin and charge density wave order as a function of $V$.
\cite{murakami00,zhang89}
We shift the chemical potential $\mu$ by $4t'$ such that for $\mu=0$ 
the Fermi surface contains the saddle points $(\pi,0)$ and $(0,\pi)$, 
where the gradient of $\epsilon_{\mathbf{k}}$ vanishes. 
The corresponding filling is referred to as Van Hove filling.

In a functional integral formalism, a one-band model of interacting
fermions is described by a bare action of the form \cite{negele87}
\begin{equation} \label{action}
 \cS[\Psi] = 
 \int \mathrm{d}k \sum_{\sigma} \ovl{\psi}_{\sigma}(k) 
 [-ik_0 + (\eps_{\bk} - \mu)] \psi_{\sigma}(k) +
 \ca{V}_0[\Psi] \; ,
\end{equation}
with anticommuting Grassmann fields $\ovl{\psi}$ and $\psi$. 
In addition to momentum $\bk$ and spin $\sigma$ the fields 
$\ovl{\psi}_{\sigma}(k)$ and $\psi_{\sigma}(k)$ also depend on 
Matsubara frequencies $k_0$.
The variable $k=(k_0,\bk)$ collects frequencies and momenta, and 
$\int \mathrm{d}k$ is a short-hand notation for 
$\int \frac{\mathrm{d}k_0}{2\pi} 
 \int \frac{\mathrm{d}^2\mathbf{k}}{(2\pi)^2}$.
We consider only the case of temperature zero in this paper, so
that the Matsubara sums are integrals.
The symbol $\Psi$ denotes a dependence on both $\psi$ and
$\ovl{\psi}$.
For the extended Hubbard model, the bare two-fermion interaction 
can be written as
\begin{align}
 \ca{V}_0[\Psi] &=
 \frac 12\int \mathrm{d} p_1 \mathrm{d} p_2 \mathrm{d} p_3 \; 
 V_0(\mathbf{p_1},\mathbf{p_2},\mathbf{p_3}) \nonumber \\ 
 & \hspace{1cm} \times \sum_{\sigma,\sigma'} 
 \ovl{\psi}_{\sigma}(p_1) \ovl{\psi}_{\sigma'} (p_2)
 \psi_{\sigma'}(p_3) \psi_{\sigma}(p_4) ,
\end{align}
where $p_4=p_1+p_2-p_3$ is fixed by momentum and frequency 
conservation, and
\begin{equation}
 V_0(\mathbf{p_1},\mathbf{p_2},\mathbf{p_3}) = 
 U +V g(\mathbf{p_2}-\mathbf{p_3}) ,
\end{equation}
with $g(\mathbf{q}) = 2 (\cos q_x +\cos q_y)$.


\section{Weak coupling functional RG}

We compute the effective two-particle interaction of the extended
Hubbard model from a functional RG flow.\cite{metzner12}
To this end we regularize the bare action by introducing a smooth
frequency cutoff $\Lambda$, corresponding to a regularized bare
propagator
\begin{equation} \label{G_0}
 G_0^{\Lambda}(k) =  
 \frac{k_0^2}{k_0^2 + \Lambda^2} \,
 \frac{1}{ik_0 - \eps_{\bk} + \mu} \, .
\end{equation}
The effective interaction $\ca{V}^{\Lam}$ on scale $\Lambda$
interpolates smoothly between the bare interaction $\ca{V}_0$ at 
$\Lambda = \infty$ and the final effective two-particle interaction
obtained in principle for $\Lambda \to 0$.
The exact flow of $\ca{V}^{\Lam}$ is determined by a differential 
flow equation, which involves effective $m$-particle interactions 
of arbitrary order $m$.\cite{metzner12}
We use a weak-coupling truncation of the exact flow equation,
where $m$-particle contributions with $m \geq 3$ and self-energy
feedback to the flow of $\ca{V}^{\Lam}$ are neglected.
Corrections to this approximation are of order $(\ca{V}^{\Lam})^3$.
The truncated flow equation has the schematic form
\begin{equation} \label{eq:FlowV}
 \frac{d}{d\Lambda} \ca{V}^{\Lambda} =
 \frac 12 {\rm tr} \left( 
 \dot G_0^{\Lambda} \frac{\partial^2 \ca{V}^{\Lambda}}{\partial\Psi^2}
 G_0^{\Lambda} \frac{\partial^2 \ca{V}^{\Lambda}}{\partial\Psi^2}
 \right) \, ,
\end{equation}
where $\dot G_0^{\Lambda} = \frac{d}{d\Lambda} G_0^{\Lambda}$.
Graphically this corresponds to two vertices connected by two 
regularized propagators (one differentiated) forming particle-particle 
or particle-hole bubbles.

We neglect the frequency dependence of the effective interaction,
which is irrelevant in power-counting at weak coupling.
To obtain an efficient parametrization of the momentum dependence
we make an ansatz \cite{husemann09}
\begin{align} \label{eq:Ansatz}
 \ca{V}^{\Lambda}[\Psi] = 
 \ca{V}_0[\Psi] + 
 \ca{V}^{\Lambda}_{\mtin{M}}[\Psi] + 
 \ca{V}^{\Lambda}_{\mtin{K}}[\Psi] + 
 \ca{V}^{\Lambda}_{\mtin{D}}[\Psi] ,
\end{align}
which is a decomposition into spin, charge, and pairing channels. 
The spin or magnetic channel describes the interaction of spin operators
\begin{align}
 \ca{V}^{\Lambda}_{\mtin{M}}[\Psi] = -
 \sum_{n=1}^2 \int \mathrm{d}q \; M_n^{\Lambda}(\mathbf{q}) \sum_{a=1}^3
 S_n^{(a)}(q) S_n^{(a)}(-q)
\end{align}
with 
$S_n^{(a)}(q) = \frac 12 \int \mathrm{d}k  \sum_{\sigma\sigma'}
 \ovl{\psi}_{\sigma}(k) \tau^{(a)}_{\sigma\sigma'}
 \psi_{\sigma'}(k+q) f_n(\mathbf{k} + \frac{\mathbf{q}}{2})$. 
Here $\tau^{(a)}$ are Pauli matrices, and $f_1(\mathbf{k}) = 1$ and 
$f_2(\mathbf{k}) = \cos k_x-\cos k_y$ are $s$-wave and $d$-wave form 
factors, respectively. 
The coupling functions $M_n^{\Lambda}(\mathbf{q})$ evolve in the RG 
flow. If $M_1^{\Lambda}(\mathbf{q})$ develops a peak at (close to)  
$\mathbf{q}=(\pi,\pi)$ which diverges upon lowering $\Lam$,
the system tends to (incommensurate) antiferromagnetic order.

A similar parametrization is chosen for the charge channel
\begin{align}
 \ca{V}^{\Lambda}_{\mtin{K}}[\Psi] = - \frac{1}{4}
 \sum_{n=1}^2 \int \mathrm{d}q \; K_n^{\Lambda}(\mathbf{q}) 
 N_n(q) N_n(-q),
\end{align}
where 
$N_n(q) = \int \mathrm{d}k \sum_{\sigma}\ovl{\psi}_{\sigma}(k)
 \psi_{\sigma}(k+q) f_n(\mathbf{k}+\frac{\mathbf{q}}{2})$ 
are density operators.
An enhancement of $K_1^{\Lambda}(\mathbf{q})$ at a finite wave 
vector $\mathbf{q}$ indicates a tendency toward formation of a
charge density wave.
In this work, we pay particular attention to the coupling function 
$K_2^{\Lambda}(\mathbf{q})$ near $\mathbf{q}=0$, which, if 
sufficiently large, can drive a $d$-wave Pomeranchuk instability 
leading to a nematic state. 
At $\mathbf{Q}=(\pi,\pi)$ the coupling $K_2(\mathbf{Q})$ will not be 
very pronounced, because for $\mathbf{k}$ sitting on a Van Hove point,
$\mathbf{k}+\frac{\mathbf{Q}}{2}$ points to a zone diagonal where 
the form factor $f_2$ is zero. 
Note that $\bra N_2(Q) \ket$ with $Q = (0,\bQ)$ differs from the
familiar $d$-density wave order parameter \cite{charkravarty01} 
$i \bra \int\mathrm{d}k \sum_{\sigma} \ovl{\psi}_{\sigma}(k)
 \psi_{\sigma}(k+Q) f_2(\mathbf{k}) \ket$. 

The pairing channel describes the interaction
\begin{align}
 \ca{V}^{\Lambda}_{\mtin{D}}[\Psi] = -
 \sum_{n=1}^2\int \mathrm{d}q \; D_n^{\Lambda}(\mathbf{q}) 
 \ovl{X}_n(q) X_n(q)
\end{align}
with spin singlet Cooper pairs
\begin{align*}
 \ovl{X}(q) &= \int \mathrm{d}k \;
 \ovl{\psi}_{\up}(k)\ovl{\psi}_{\down}(q-k) 
 f_n(\sfrac{\bq}{2} - \bk) \, , \\
X(q) &= \int \mathrm{d}k \;\psi_{\down}(k)\psi_{\up}(q-k) 
 f_n(\sfrac{\bq}{2} - \bk) \, . 
\end{align*}
The form factors determine the symmetry of the gap. A diverging
(upon lowering $\Lam$) peak of $D_1^{\Lambda}(\mathbf{q})$ at 
$\mathbf{q} = 0$ indicates an $s$-wave pairing instability, and 
a diverging peak of $D_2^{\Lambda}(\mathbf{q})$ at $\mathbf{q}=0$ 
a $d$-wave pairing instability, so that a superconducting state
is favored. 
Since both form factors are even functions of momentum we only 
consider singlets in the spin dependence.

The restriction to $s$-wave and $d$-wave form factors in our
ansatz is motivated by the dominance of $s$-wave and $d$-wave
interactions observed in all previous fRG studies of Hubbard-type 
models in the parameter range considered here.\cite{metzner12}

Substituting the ansatz Eq.~(\ref{eq:Ansatz}) into the flow 
equation generates a variety of different one-loop graphs, 
not all being of the form of the ansatz. 
We will use a suitable approximate projection which was developed 
in Ref.~\onlinecite{husemann09}. 
First, the graphs are sorted according to the transfer momentum that 
propagates through the fermion loop. As long as the coupling functions 
do not have pronounced peaks this is the main momentum dependence and 
the graphs can be assigned to one channel. Since the form factors $f_1$ 
and $f_2$ are orthonormal functions on $[-\pi,\pi]^2$, the graphs in 
each channel can be projected to the flow of coupling functions by 
integration. The resulting flow equations can be found in 
Ref.~\onlinecite{husemann09} and have been shown to yield a reasonable 
approximation to the RG flow.

The above truncation of the RG is a weak coupling approximation. 
For high scales, $U/\Lambda$ is a small parameter, which allows 
to truncate at one-loop order. Generically couplings start to grow 
when the scale is descreased. In an intermediate scale regime there 
are phase space arguments such that the one-loop flow remains valid 
although the couplings are not small anymore.\cite{salmhofer01} 
Eventually, the coupling functions become too large for certain 
momenta and the truncation becomes unreliable. 
We then stop the flow and interpret the rapid growth of the leading
coupling as an instability indicating a corresponding ordered state. 
Specifically, we stop the flow when the maximum of the coupling 
functions exceeds $V_{\max}=20t$.
The stopping scale $\Lambda_*$ is an upper bound for the critical 
scale, at which interactions diverge.


\section{Effective d-wave Interactions}

We now present results for the effective $d$-wave interactions in the
charge and pairing channel in the presence of strong antiferromagnetic
correlations.
Throughout this section we choose a relatively weak Hubbard 
interaction $U = 3t$.

The leading instabilities are signalled by diverging coupling
functions in the corresponding channel. The divergence of the
full two-particle vertex is well captured by these coupling
functions.
However, to compare effective interactions quantitatively,
one has to sum the contributions from all terms in the 
decomposition Eq.~(\ref{eq:Ansatz}), including also the bare
interaction $\ca{V}_0$.
Contributions from the bare interactions and other channels
can be significant in channels where the effective interaction 
remains bounded.

The complete effective interaction in the various channels 
can be extracted from the two-particle vertex
$\Gamma_{\sg_1\sg_2;\sg_3\sg_4}^{\Lam}(p_1,p_2;p_3,p_4)$,
which is the antisymmetrized kernel of $\ca{V}^{\Lam}$,
that is,
\begin{eqnarray}
 \ca{V}^{\Lam}[\Psi] &=& 
 \frac{1}{4} \sum_{\sg_1,\dots,\sg_4} 
 \int \! \mathrm{d} p_1 \mathrm{d} p_2 \mathrm{d} p_3 \,
 \Gamma_{\sg_1\sg_2;\sg_3\sg_4}^{\Lam}(p_1,p_2;p_3,p_4)
 \nonumber \\
 &\times& 
 \ovl{\psi}_{\sg_1}(p_1) \ovl{\psi}_{\sg_2}(p_2)
 \psi_{\sg_3}(p_3) \psi_{\sg_4}(p_4) \, ,
\end{eqnarray}
where $p_4 = p_1 + p_2 - p_3$.
We denote the static limit of $\Gamma^{\Lam}$ by
$\Gamma_{\sg_1\sg_2;\sg_3\sg_4}^{\Lam}(\bp_1,\bp_2;\bp_3,\bp_4)$.
The effective singlet pairing interaction is then given by
\begin{equation}
 \Gam_{\bp\bp'}^{\mtin{SC},\Lam} = 
 \Gam_{\up\down;\down\up}^{\Lam}(\bp,-\bp;-\bp',\bp') \, ,
\end{equation}
and the effective charge interaction by
\begin{equation}
 \Gam_{\bp\bp'}^{\mtin{C},\Lam}(\bq) = 
 \frac{1}{4} \sum_{\sg,\sg'} 
 \Gam_{\sg\sg';\sg'\sg}^{\Lam}(\bp,\bp';\bp'-\bq,\bp+\bq) \, ,
\end{equation}
where $\bq$ is the momentum transfer.
In a renormalized mean-field theory of spin singlet 
superconductivity based on $\Gamma^{\Lam}$, only the pairing 
interaction $\Gam_{\bp\bp'}^{\mtin{SC},\Lam}$ contributes.
\cite{reiss07}
Similarly, a renormalized mean-field theory for charge order
with a modulation vector $\bq$ involves exclusively the charge
interaction $\Gam_{\bp\bp'}^{\mtin{C},\Lam}(\bq)$.

The s-wave and d-wave components of the interactions are
extracted by the projections
\begin{equation} \label{eq:gamSC}
 \gam_n^{\mtin{SC},\Lam} = \int
 \frac{\mathrm{d}^2\bp}{(2\pi)^2} 
 \frac{\mathrm{d}^2\bp'}{(2\pi)^2} \,
 f_n(\bp) f_n(\bp') \, \Gam_{\bp\bp'}^{\mtin{SC},\Lam}
\end{equation}
and
\begin{equation} \label{eq:gamC}
 \gam_n^{\mtin{C},\Lam}(\bq) = \int
 \frac{\mathrm{d}^2\bp}{(2\pi)^2} 
 \frac{\mathrm{d}^2\bp'}{(2\pi)^2} \,
 f_n(\bp+\sfrac{\bq}{2}) f_n(\bp'-\sfrac{\bq}{2}) \, 
 \Gam_{\bp\bp'}^{\mtin{C},\Lam}(\bq)
\end{equation}
with $n = 1,2$.
Inserting the channel decomposition Eq.~(\ref{eq:Ansatz}), the
$d$-wave components can be written as
\begin{equation}
 \gam_2^{\mtin{SC},\Lam} = V - D_2^{\Lam}(\b0) + \dots \, ,
\end{equation}
and
\begin{equation}
 \gam_2^{\mtin{C},\Lam}(\bq) = 
 - \frac{V}{2} - \frac{1}{2} K_2^{\Lam}(\bq) + \dots \, ,
\end{equation}
where the dots refer to fluctuation contributions from
other channels.
Complete expressions for $\gam_n^{\mtin{SC},\Lam}$
and $\gam_n^{\mtin{C},\Lam}(\bq)$ in terms of the bare
interaction and the coupling functions $M_n^{\Lam}$,
$K_n^{\Lam}$ and $D_n^{\Lam}$ are given in the Appendix. 
Note that positive coupling functions yield attractive
(negative) contributions to the corresponding effective 
interactions.

\subsection{Van Hove filling}

We first consider the case of Van Hove filling for various choices
of $t' < 0$. 
For $|t'| < 0.237t$ the RG flow runs into an antiferromagnetic
instability, that is, $M_1^{\Lambda}(\mathbf{q})$ with $\mathbf{q}$ 
close to $(\pi,\pi)$ reaches $V_{\max}$ first.
For $|t'| < 0.11t$ the maximum of $M_1^{\Lambda_*}(\mathbf{q})$ 
is situated at $\bQ = (\pi,\pi)$, while it deviates from $(\pi,\pi)$ 
for larger $|t'|$.
However, this shift of $\bq$ depends sensitively on the choice
of $\Lambda_*$, and the maximal value of $M_1^{\Lambda_*}(\mathbf{q})$
differs only very little from its value at $\bQ$.
For $|t'| > 0.237t$ the dominant instability is $d$-wave pairing,
that is, $D_2^{\Lambda}(\mathbf{0})$ is the largest coupling at 
$\Lambda_*$.
The transition from antiferromagnetism to $d$-wave pairing upon
increasing $|t'|$ was found already in the first fRG studies of
the two-dimensional Hubbard model.\cite{halboth00,honerkamp01}

In Fig.~1 we show results for the effective $d$-wave interactions
at the scale $\Lambda_*$ as a function of $t'/t$.
The antiferromagnetic (AF) and $d$-wave pairing (dSC) regions are 
separated by a vertical dashed line.
The stopping scale $\Lam_*$ decreases monotonically from $0.194t$
at $t' = 0$ to $0.013t$ at $t' = -0.28t$ for $V = 0$, and from 
$0.162t$ at $t' = 0$ to $0.007t$ at $t' = -0.28t$ for $V = 0.74t$.
\begin{figure}
\begin{center}
\includegraphics[width=0.35\textwidth,angle=0]{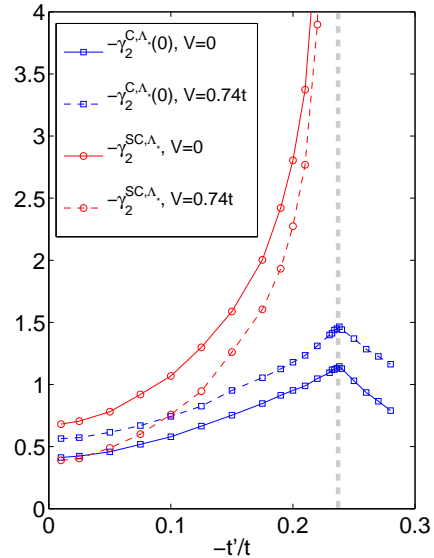}
\caption{(Color online) Effective $d$-wave interactions 
 at the scale $\Lambda_*$ in units of $t$ as a function of $t'$,
 for $U=3t$ at Van Hove filling. The electron density ranges 
 from $n=1$ at $t'=0$ to $n=0.75$ at $t'/t = -0.28$.
 The pairing interactions $\gamma_2^{\mtin{SC},\Lambda_*}$ are 
 plotted as circles and charge interactions 
 $\gamma_2^{\mtin{C},\Lambda_*}(\mathbf{0})$ as squares. 
 Solid lines correspond to $V=0$ and dashed lines to $V=0.74t$. 
 In both cases the leading instability changes from
 antiferromagnetism to $d$-wave pairing at $t'=-0.237t$
(corresponding to $n=0.79$).}
\end{center}
\end{figure}
At Van Hove filling, the $d$-wave charge interaction 
$\gamma_2^{\mtin{C},\Lambda_*}(\bq)$ is peaked at $\bq = \mathbf{0}$.
For the plain Hubbard model ($V = 0$) it is always weaker than
the $d$-wave pairing interaction $\gamma_2^{\mtin{SC},\Lambda_*}$, 
as observed already in Ref.~\onlinecite{honerkamp02}.
A finite $V$ enhances the $d$-wave charge interaction, but it 
remains small compared to the dominant antiferromagnetic and
pairing interactions at small and large $|t'|$, respectively.

In Fig.~2 we plot the effective interactions 
$\gamma_2^{\mtin{C},\Lambda_*}(\mathbf{0})$ and $\gamma_2^{\mtin{SC},\Lambda_*}$ 
as functions of $V$ at $t' = -0.15t$.
From $V = 0$ to $V = 0.758t$ the stopping scale decreases moderately 
from $\Lambda_* = 0.078t$ to $\Lambda_* = 0.053t$, but it then increases
rapidly to $\Lambda_* = 0.656t$ at $V = t$. 
\begin{figure}
\begin{center}
\includegraphics[width=0.35\textwidth,angle=0]{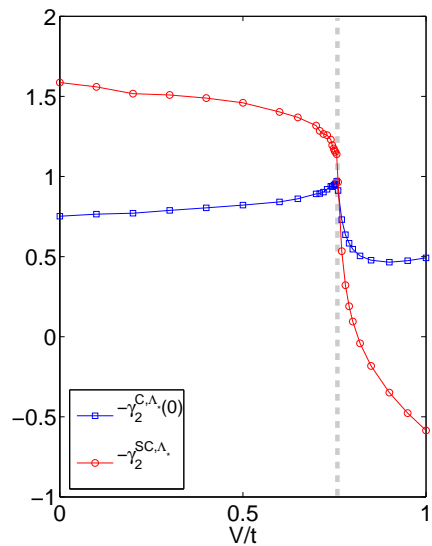}
\caption{(Color online) Effective $d$-wave interactions
 $\gamma_2^{\mtin{SC},\Lambda_*}$ (circles)
 and $\gamma_2^{\mtin{C},\Lambda_*}(\mathbf{0})$ (squares) in units of $t$ as a
 function of the nearest-neighbor interaction $V$, for $t' = -0.15t$ and
 $U=3t$ at Van Hove filling ($n=0.88$). 
 The leading instability changes from incommensurate antiferromagnetism 
 to charge density wave at $V = 0.758t$.}
\end{center}
\end{figure}
For increasing $V < 0.758t$, in terms of absolute values, the $d$-wave
interaction in the density channel increases while the $d$-wave pairing
interaction decreases. 
For $V > 0.758t$ the leading instability changes from incommensurate
antiferromagnetism to a charge density wave, that is, the coupling 
function $K_1^{\Lambda}(\bQ)$ reaches $V_{\max}$ first and defines 
the stopping scale $\Lambda_*$.
The critical value $V = 0.758t$ is close to the mean-field 
transition point \cite{murakami00} at $V = U/4 = 0.75t$.
The drop of $-\gamma_2^{\mtin{C},\Lambda_*}(\mathbf{0})$ and
$-\gamma_2^{\mtin{SC},\Lambda_*}$ 
for $V > 0.758t$ is due to a rapidly increasing $\Lambda_*$ (for
increasing $V$) in the charge density wave regime.

\subsection{Away from Van Hove filling}

In the following we fix the next-to-nearest neighbor hopping amplitude 
to $t'= - 0.15t$ and vary the chemical potential $\mu$ in a limited
range around Van Hove filling.
For this choice of parameters the RG flow runs into an antiferromagnetic
instability, that is, $M_1^{\Lambda}(\mathbf{q})$ with $\mathbf{q}$ 
close to $(\pi,\pi)$ reaches $V_{\max}$ first. 
For $\mu > 0.03t$ the maximum of $M_1^{\Lambda_*}(\mathbf{q})$ is at 
$\bQ = (\pi,\pi)$, and for $\mu < 0.03t$ four maxima are found at 
$(\pi,\pi\pm \delta)$ and $(\pi\pm \delta,\pi)$, with $\delta$ increasing 
upon lowering $\mu$. 
The dominance of antiferromagnetic correlations is not affected by a 
nearest neighbor interaction $V$ as long as $V < U/4$.
At sufficiently low $\mu$ (large hole doping), $d$-wave pairing 
becomes the leading instability.\cite{zanchi00,halboth00,honerkamp01}
Here we do not enter the $d$-wave superconducting region.

Results for the effective $d$-wave interactions $\gamma_2^{\mtin{SC},\Lambda_*}$
and 
$\gamma_2^{\mtin{C},\Lambda_*}$ in the antiferromagnetic background are shown
for the plain Hubbard model ($V=0$) in Fig.~3.
\begin{figure}
\begin{center}
\includegraphics[width=0.35\textwidth,angle=0]{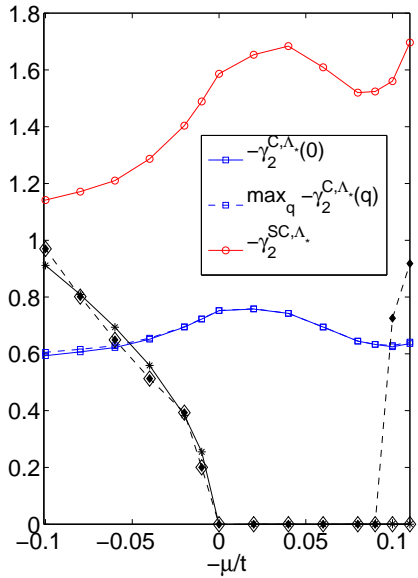}
\caption{(Color online) Effective $d$-wave interactions at 
the stopping scale $\Lambda_*$ of antiferromagnetism plotted as functions 
of $\mu$ for $t'=-0.15t$, $U=3t$, and $V=0$.
The density varies from $n=0.94$ at $\mu/t = 0.1$ to $n=0.81$ at
$\mu/t= -0.11$. 
Pairing interactions $\gamma_2^{\mtin{SC},\Lambda_*}$ are plotted as
circles connected by a solid line. The most attractive $d$-wave charge
interactions $\max_{\mathbf{q}} |\gamma_2^{\mtin{C},\Lambda_*}(\mathbf{q})|$ 
are plotted as squares
connected by a dashed line, while the points corresponding to 
$\gamma_2^{\mtin{C},\Lambda_*}({\bf 0})$ are connected by a solid line.
Also shown are momenta $\mathbf{q}$ where
$-\gamma_2^{\mtin{C},\Lambda_*}(\mathbf{q})$ 
becomes maximal, with components $q_x$ plotted as open diamonds and
$q_y$ as
filled diamonds, both connected by a dashed line.
Momenta for which the bare $d$-wave particle-hole bubble
$\Phi_{\mtin{ph}}^{2,\Lambda_*}(\mathbf{q})$ is extremal are 
shown similarly with components $q_x$ plotted as plus signs and $q_y$ as
diagonal crosses. Stars are superpositions of plus signs and crosses for $q_x =
q_y$.}
\end{center}
\end{figure}
We distinguish between
$\gamma_2^{\mtin{SC},\Lambda_*}(\mathbf{0})$ (solid line) 
and $\max_{\bq} |\gamma_2^{\mtin{C},\Lambda_*}(\bq)|$ (dashed line). 
Above Van Hove filling ($\mu > 0$) there are four maxima of 
$-\gamma_2^{\mtin{C},\Lambda_*}(\bq)$ at $\bq = (\pm\delta_1,\pm\delta_1)$ 
with $\delta_1$ decreasing as we approach Van Hove filling. 
The positions of these maxima are plotted in Fig.~3 as well.
They are very close to the positions of the maxima of the 
static regularized $d$-wave particle-hole bubble
\begin{align}\label{eq:phbubble}
 \Phi^{2,\Lambda}_{\mtin{ph}}(\mathbf{q}) = 
 \int \mathrm{d} p \; 
 G_0^{\Lambda}(p_0,\bp) G_0^{\Lambda}(p_0,\bp+\bq)
 f_2(\bp + \sfrac{\bq}{2})^2 
\end{align}
at $\Lambda = \Lambda_*$, which are plotted in Fig.~3 for comparison.
The small deviations are not significant, that is, they may
be due to the limited momentum resolution in the numerics.

A large attraction $\gamma_2^{\mtin{C},\Lambda_*}(\bq)$ at a wave 
vector $\bq \neq \mathbf{0}$ would signal a tendency to form a 
modulated nematic order with a modulation vector $\bq$.
Such a tendency was found in a recent analysis of secondary
instabilities generated by antiferromagnetic fluctuations 
in a two-dimensional metal by Metlitski and Sachdev.
\cite{metlitski10a,metlitski10b}
The modulation vectors $\bq$ are fixed by the position of 
the antiferromagnetic hot spots, where the antiferromagnetic
fluctuations couple most strongly to the electronic 
excitations near the Fermi surface.
Subsequently it was shown that the same modulation vectors
$\bq = (\pm\delta_1,\pm\delta_1)$ emerge as the momenta where 
the static $d$-wave particle-hole bubble is maximal.
\cite{holder12}
For a tight-binding dispersion with hopping amplitudes $t$ and
$t'$, the modulation is given by \cite{holder12}
\begin{equation} \label{delta_1}
 \delta_1 = 2\arccos\sqrt{1 - \frac{\mu}{4|t'|}} \, .
\end{equation}
Note that $\mu \geq 0$ has been shifted by $4t'$ such that 
Van Hove filling corresponds to $\mu = 0$.
The maxima of the $d$-wave interaction $-\gamma_2^{\mtin{C},\Lambda_*}(\bq)$  
obtained from the fRG flow above Van Hove filling are also 
situated at diagonal momenta of the form 
$\bq = (\pm\delta_1,\pm\delta_1)$, with $\delta_1$ given by 
Eq.~(\ref{delta_1}) within the numerical resolution. 
In agreement with Metlitski and Sachdev 
\cite{metlitski10a,metlitski10b} we find that in the 
antiferromagnetic regime the $d$-wave density interaction is of 
the same order of magnitude but smaller than the $d$-wave 
pairing interaction.
However, while $d$-wave pairing competes with antiferromagnetism
at larger doping, the $d$-wave density interaction remains rather
small.

Slightly below Van Hove filling the particle-hole bubble is 
very flat. 
In the numerical solution of the RG flow we find the maximum of 
$-\gamma_2^{\mtin{C},\Lambda_*}(\bq)$ at $\bq = 0$ for $-0.08t < \mu < 0$. 
For lower electron filling, $\mu < -0.08t$, there are again four 
degenerate maxima, but now at momenta along the crystal axes
$(0,\pm\delta_2)$ and $(\pm\delta_2,0)$. 
Overall, the strength of the effective $d$-wave interaction in the density 
channel is weaker than in the $d$-wave pairing channel and is 
maximal around Van Hove filling.

A nearest neighbor repulsion is expected to enhance the 
$d$-wave charge correlations. 
Indeed, in a mean-field study restricted to spatially 
homogeneous solutions it was found that the nearest
neighbor interaction in the extended Hubbard model can
generate a nematic state.\cite{valenzuela01}

In Fig.~4 we show results for the effective $d$-wave interactions for
the extended Hubbard model with a finite nearest-neighbor interaction
$V = 0.74t$. The other parameters are the same as in Fig.~3.
\begin{figure}
\begin{center}
\includegraphics[width=0.35\textwidth,angle=0]{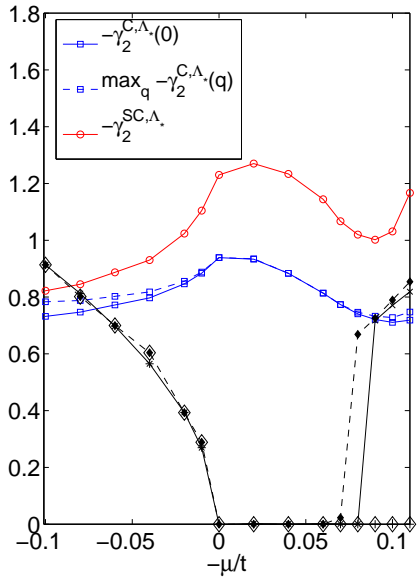}
\caption{(Color online) Effective $d$-wave interactions 
 and modulation vectors as in Fig.~3, but for a finite nearest 
 neighbor interaction $V = 0.74t$.}
\end{center}
\end{figure}
The $d$-wave density interaction is enhanced by the presence 
of $V$, while the pairing interaction is reduced compared 
to the case $V=0$. The favorite nematic modulation vectors 
are not affected significantly by $V$, but the peaks of 
$\gamma_2^{\mtin{C},\Lambda_*}(\bq)$ at those wave vectors are 
more pronounced in Fig.~4 compared to the case $V=0$ plotted 
in Fig.~3.
The stopping scale $\Lambda_*$ decreases monotonically from
$0.105t$ at $\mu/t = 0.1$ to $0.036t$ at $\mu/t = - 0.11$ for $V = 0$
(Fig.~3), and from $0.076t$ at $\mu/t = 0.1$ to $0.019t$ at 
$\mu/t = - 0.11$ for $V = 0.74t$ (Fig.~4).

We focused on the case of small or moderate doping.
Further away from half-filling, modulated {\em ferromagnetic} 
fluctuations with incommensurate wave vectors have been found
in mean-field and functional RG studies of the two-dimensional
Hubbard model.\cite{igoshev10,igoshev11,katanin11}

\subsection{Mechanism}

To gain some analytical understanding of the above results, 
we now discuss the structure of the flow equations for the 
effective $d$-wave couplings.
The flow equation for the $d$-wave coupling in the density 
channel can be written as \cite{husemann09}
\begin{align} \label{eq:dotK2}
 \frac{d}{d\Lambda} K_2^{\Lambda}(\bq) = 
 - \frac{d\Phi_{\mtin{ph}}^{2,\Lambda}(\bq)}{d\Lambda} 
 \left[ K_2^\Lambda(\bq) + V - \alpha_K^{\Lambda} \right ]^2 
 + F_K^{\Lambda}(\bq) ,
\end{align}
where $\alpha_K^{\Lambda}$ is a contribution from $s$-wave 
channels given by
\begin{eqnarray} \label{eq:alphaK}
 \alpha_{K}^{\Lambda} &=& 
 \frac{1}{4} \int\frac{\mathrm{d}^2\bk}{(2\pi)^2} 
 \left( \cos k_x + \cos k_y \right)
 \nonumber \\[2mm]
 && \times \left[ -2 D_1^{\Lambda}(\bk) + 3M_1^{\Lambda}(\bk) 
 + K_1^{\Lambda}(\bk) \right] \, .
\end{eqnarray}
The function $F_K^{\Lambda}(\bq)$ originates from $d$-wave 
couplings $D_2^{\Lambda}$, $M_2^{\Lambda}$, and $K_2^{\Lambda}$. 
We do not write this function here,\cite{husemann09} but it
was taken into account in our numerical solution of the flow
equations.
As long as the $d$-wave interactions are small, contributions
from $F_K^{\Lambda}(\mathbf{q})$ are negligible. 
Then the only momentum dependence is generated by a scale 
derivative of the regularized $d$-wave particle-hole bubble 
Eq.~(\ref{eq:phbubble}). 
The initial condition at a high scale $\Lambda_0$, derived 
from second order perturbation theory, is also given by this 
bubble $K_2^{\Lambda_0}(\bq) = 
- V^2 \Phi_{\mtin{ph}}^{2,\Lambda_0}(\bq) > 0$. 
This makes plausible why the maxima of $K_2^{\Lambda}(\bq)$ and
$-\gamma_2^{\mtin{C},\Lambda}(\bq)$ 
are given by the extrema of the regularized particle-hole bubble.

Even if there is no $d$-wave coupling initially (for $V=0$), 
$K_2^{\Lambda}$ is generated in the flow similarly to the 
mechanism that generates an attraction leading to $d$-wave 
pairing. 
Close to an antiferromagnetic instability the magnetic coupling 
function obeys $M_1^{\Lambda}(\bQ) \gg M_1^{\Lambda}(\mathbf{0})$ 
with $\bQ = (\pi,\pi)$. Via integration with the cosine this leads 
to a sizeable negative contribution in Eq.~(\ref{eq:alphaK}). 
Then the square bracket in Eq.~(\ref{eq:dotK2}) remains non-zero 
for $K_2^{\Lambda} > 0$ and increases with the latter. 
Since $\partial_{\Lambda}\Phi_{\mtin{ph}}^{2,\Lambda}(\bq) > 0$ 
and the scale $\Lambda$ decreases, the coupling function 
$K_2^{\Lambda}(\bq)$ builds up in the flow.
It is not essential whether $M_1^{\Lambda}(\bk)$ is maximal at
or close to $(\pi,\pi)$. The modulation of the $d$-wave coupling
is not tied to an incommensurability in the magnetic interactions.

In the same way $K_2^{\Lambda}$ is generated by the density 
channel if $K_1^{\Lambda}(\bQ) \gg K_1^{\Lambda}(\mathbf{0})$, 
that is, if the system is close to a charge density wave 
instability (CDW). The latter is the case for larger $V$, which 
also helps to generate $K_2^{\Lambda}$ via the constant $V$ in 
Eq.~(\ref{eq:dotK2}) and via a larger initial condition. 
Therefore we find a larger effective $d$-wave interaction in 
the density channel if the nearest neighbor interaction $V$ is 
increased, as in Figs.~2 and 4.

The flow equation for $D_2^{\Lambda}$ is given by \cite{husemann09}
\begin{align} \label{eq:dotD2}
 \frac{d}{d\Lambda} D_2^{\Lambda}(\bq) = 
 \frac{d\Phi_{\mtin{pp}}^{2,\Lambda}(\bq)}{d\Lambda} 
 \left[ D_2^{\Lambda}(\bq) - V - \alpha_D^{\Lambda} \right]^2 +
 F_D^{\Lambda}(\bq) \, ,
\end{align}
where $\Phi_{\mtin{pp}}^{2,\Lambda}(\bq) = \int \mathrm{d}p \, 
 G_0^{\Lambda}(p_0,\bp) G_0^{\Lambda}(-p_0,\bq-\bp) 
 f_2(\frac{\bq}{2}-\bp)^2$ 
is the static regularized particle-particle bubble with $d$-wave 
form factors and
\begin{align} \label{eq:alphaD}
 \alpha_D^{\Lambda} = 
 \frac{1}{4} \int \frac{\mathrm{d}^2\bk}{(2\pi)^2} 
 \left( \cos k_x + \cos k_y \right)  
 \left[ 3M_1^{\Lambda}(\bk) - K_1^{\Lambda}(\bk) \right] \, .
\end{align}
Again we neglect the function $F_D^{\Lambda}(\bq)$ for simplicity, 
which is justified for small $K_2^{\Lambda}$ and $M_2^{\Lambda}$. 
The initial condition at a high scale $\Lambda_0$, obtained from 
second order perturbation theory, is given by 
$D_2^{\Lambda_0}(\bq) = V^2 \Phi_{\mtin{pp}}^{2,\Lambda}(\bq) > 0$ 
and is dominated by the constant $-V$ in Eq.~(\ref{eq:dotD2}). 
Because $\partial_{\Lambda}\Phi_{\mtin{pp}}^{2,\Lambda}(\mathbf{0}) < 0$ 
and $\Lambda$ decreases, the coupling $D_2^{\Lambda}(\mathbf{0})$ would 
saturate at $V$ if $\alpha_D^{\Lambda} = 0$. 
The $d$-wave pairing can only become large if $\alpha_D^{\Lambda}$ 
is negative. Like for $K_2^{\Lambda}$ this is the case in an 
antiferromagnetic background, which is the main mechanism for 
$d$-wave superconductivity in the Hubbard model. 
However, because of the minus sign in front of $K_1^{\Lambda}(\bk)$
in Eq.~(\ref{eq:alphaD}), the vicinity of charge density order is 
counterproductive for the evolution of $d$-wave pairing.

It is instructive to compare the contribution from magnetic 
interactions to the generation of $d$-wave couplings described
above to the hot spot mechanism discovered by Metlitski and 
Sachdev.\cite{metlitski10a,metlitski10b} 
They consider the situation at a quantum critical point on the 
phase boundary of a commensurate antiferromagnetic ground state. 
As a consequence, their spin fluctuation propagator is strongly
peaked at $(\pi,\pi)$, and electronic excitations around hot 
spots dominate. The nematic modulation vector is determined by 
the distance between hot spots with collinear Fermi velocities.
Metlitski and Sachdev find a degeneracy between the $d$-wave 
couplings in the pairing and charge channels, which is broken
only by the Fermi surface curvature. This degeneracy is present 
also in our flow equations Eqs.\ (\ref{eq:alphaK}) and 
(\ref{eq:alphaD}), where $M_1^{\Lambda}(\bk)$ contributes
exactly equally to $\alpha_K^{\Lambda}$ and $\alpha_D^{\Lambda}$.
The Fermi surface curvature lifts the degeneracy because it
reduces the size of the particle-hole bubble at the nematic 
modulation vector compared to the particle-particle bubble at 
$\bq = \b0$.
Singular self-energy contributions at the quantum critical 
point suppress the curvature effects, so that the nematic
correlations are more pronounced than away from criticality.
\cite{metlitski10a}

\section{Conclusion}

We computed effective $d$-wave interactions in the two-dimensional
extended Hubbard model at small to moderate distance from half-filling
by using a functional renormalization group flow.
In addition to the well-known attraction in the $d$-wave pairing
channel, an attractive interaction in the $d$-wave charge channel is 
generated, in agreement with early fRG calculations.\cite{halboth00a}
If strong enough, that interaction could induce a $d$-wave Pomeranchuk
instability leading to nematic order.
However, in comparison to the dominating antiferromagnetism and 
$d$-wave pairing, the $d$-wave charge interaction is found to be 
relatively weak. 
It becomes most pronounced at Van Hove filling and can be 
enhanced by a nearest neighbor interaction $V$. 

The $d$-wave charge attraction is not necessarily maximal at
$\bq = 0$. 
Above Van Hove filling, we find four degenerate peaks of
$\gamma_2^{\mtin{C},\Lambda}(\bq)$ at diagonal wave vectors 
$\bq = (\pm\delta_1, \pm\delta_1)$ connecting hot spots with 
collinear Fermi velocities, corresponding to the modulated nematic 
correlations discovered by Metlitski and Sachdev.
\cite{metlitski10a,metlitski10b}
In the fRG calculation, the peaks are essentially determined
by the structure of the $d$-wave particle-hole bubble, discussed
in detail in Ref.~\onlinecite{holder12}.
Below but close to Van Hove filling $|\gamma_2^{\mtin{C},\Lambda}(\bq)|$
is quite flat with a shallow maximum at $\bq = 0$. 
Further decreasing $\mu$ one obtains maxima along the crystal
axes, again corresponding to extrema in the $d$-wave bubble.

In any case, the nematic fluctuations in the weakly interacting 
extended Hubbard model seem to be generically small compared to 
the dominant channels, which, depending on parameters, are 
antiferromagnetic, $d$-wave pairing, or charge density fluctuations 
at small or moderate doping.
The nematic tendency observed experimentally in cuprates is thus 
probably a strong coupling phenomenon,\cite{kivelson98,okamoto10}
possibly associated with antiferromagnetic quantum criticality.
\cite{metlitski10a,metlitski10b} 
For a microscopic theory of the nematic order in $\rm Sr_3 Ru_2 O_7$ 
a multi-band model \cite{raghu09} seems to be required.


\begin{acknowledgments}
We are grateful to T.~Holder, A.~Katanin, M.~Metlitski, and H.~Yamase 
for valuable discussions.
\end{acknowledgments}


\begin{appendix}

\section{Expressions for effective interactions}

In this appendix we present explicit expressions for the $s$-wave and
$d$-wave components of the effective charge and pairing interactions
in terms of the bare interaction and the coupling functions.

The $s$-wave component of the effective interaction in the singlet pairing 
channel as defined in Eq.~(\ref{eq:gamSC}) can be written as
\begin{align}
 \gamma_1^{\mtin{SC},\Lambda} &= 
 U-D_1^{\Lambda}(\mathbf{0})  \\ \nonumber & \quad 
 + \frac 12 \sum_{n=1,2}\int \frac{\mathrm{d}^2\mathbf{k}}{(2\pi)^2} \; 
 \Big[ 3M^{\Lambda}_n(\mathbf{k}) -K^{\Lambda}_n(\mathbf{k}) \Big]\, .
\end{align}
Similarly, the $d$-wave component is given by
\begin{align}
 \gamma_2^{\mtin{SC},\Lambda} &= 
 V-D^{\Lambda}_2(\mathbf{0}) + \alpha_D^{\Lambda}  \\ \nonumber & \quad  
 + \frac 12 \int \frac{\mathrm{d}^2\mathbf{k}}{(2\pi)^2} \Big[\sfrac 14 
 + h_1(\mathbf{k}) + h_2(\mathbf{k}) \Big] \\ \nonumber & \qquad\qquad 
 \times  \Big[ 3M^{\Lambda}_2(\mathbf{k}) -K^{\Lambda}_2(\mathbf{k})\Big] 
 \, ,
\end{align}
with $\alpha_D^{\Lambda}$ from Eq.~(\ref{eq:alphaD}). 
The functions $h_n$ are defined as 
$h_1(\mathbf{k}) = \frac 12 (\cos k_x +\cos k_y)$ and 
$h_2(\mathbf{k}) = \cos \frac{k_x}{2} \cos \frac{k_y}{2}$. 
The first line is the main contribution and its square enters the flow 
equation of the coupling function $D_2^{\Lambda}$ in Eq.~(\ref{eq:dotD2}).

In the charge channel we allow for a finite transfer momentum $\mathbf{q}$ 
in the effective interaction between density pairs. Its $s$-wave component 
reads 
\begin{align}
 \gamma^{\mtin{C},\Lambda}_1(\mathbf{q}) &= 
 \frac 12U +V g(\mathbf{q}) - \frac 12 K^{\Lambda}_1(\mathbf{q}) + 
 a^{\mtin{C},\Lambda}_{1,1} + a^{\mtin{C},\Lambda}_{1,2} h_1(\mathbf{q})\, ,
\end{align}
where
\begin{align}
 a^{\mtin{C},\Lambda}_{1,n} &= 
 \frac 14 \int \frac{\mathrm{d}^2\mathbf{k}}{(2\pi)^2} \; 
 \Big[ -2D^{\Lambda}_n(\mathbf{k}) + 3M^{\Lambda}_n(\mathbf{k}) + 
 K^{\Lambda}_n(\mathbf{k}) \Big] \, .
\end{align}
The $d$-wave component is given by
\begin{align}
 \gamma^{\mtin{C},\Lambda}_2(\mathbf{q}) &= 
 - \frac 12 V - \frac 12 K^{\Lambda}_2(\mathbf{q}) + 
 \frac 12 \alpha^{\Lambda}_K  + A^{\mtin{C},\Lambda}_2(\mathbf{q})\, ,
\end{align}
with $\alpha_K^{\Lambda}$ from Eq.~(\ref{eq:alphaK}). 
The first three terms enter the flow equation for $K_2^{\Lambda}$, 
see Eq.~(\ref{eq:dotK2}). 
The remaining terms are subleading and are given by
\begin{align}
 A^{\mtin{C},\Lambda}_2(\mathbf{q}) &= 
 a^{\mtin{C},\Lambda}_{2,1} h_1(\mathbf{q}) + 
 a^{\mtin{C},\Lambda}_{2,2} \cos\frac{q_x}{2}\cos \frac{q_y}{2} + 
 \frac {1}{16} a^{\mtin{C},\Lambda}_{1,2} 
\end{align}
with
\begin{align}
 a^{\mtin{C},\Lambda}_{2,n} &= 
 \frac 14 \int \frac{\mathrm{d}^2\mathbf{k}}{(2\pi)^2} \; h_n(\mathbf{k}) 
 \; \\ \nonumber  & \qquad 
 \times \Big[ -2D^{\Lambda}_2(\mathbf{k}) + 3M^{\Lambda}_2(\mathbf{k}) + 
 K^{\Lambda}_2(\mathbf{k}) \Big] \, .
\end{align}

\end{appendix}


\end{document}